\documentclass[aps,preprint,epsfig]{revtex4}
\usepackage[dvips]{graphicx}
\usepackage{dcolumn}
\usepackage{bm}
\begin{document}

\draft


\title{Comparisons of Model Predictions of Pion Electrioproduction
Amplitudes in the low $Q^2$ Region}


\author{
T. Sato$^{a}$ 
and T.-S. H. Lee\,$^b$%
}


\address{
$^a$
Department of Physics, Osaka University, Toyonaka, Osaka 560-0043,
 Japan\\ 
$^b$
Physics Division, Argonne National Laboratory, Argonne,
Illinois 60439}


\begin{abstract}
Pion electroproduction amplitudes predicted by three theoretical models
in the $Q^2 \leq 0.5$ (GeV/c)$^2$ region are compared. The objective is
to facilitate the analyses of the data from new experiments on investigating
the pion cloud effects on the $\gamma N \rightarrow \Delta$ transition
form factors.
\end{abstract}


\maketitle

In recent years, the data of
pion electroproduction reactions have been analyzed by using
various models for extracting the $\gamma N \rightarrow
\Delta$ transition form factors. The large pion cloud effects predicted
by the dynamical models\cite{satolee1,satolee2,dmt1,dmt2} have motivated new
 measurements in the low mometum transfer region $Q^2 \leq 0.5 $
(GeV/c)$^2$. To facilitate the analyses of these new data, we here
compare the predictions from three models : the Sato-Lee(SL) 
Model\cite{satolee1,satolee2}, the Dubna-Mainz-Taiwan(DMT) 
model\cite{dmt1,dmt2}, 
and the Mainz Unitary Isobar Model (MAID)\cite{maid}. 
The details of each of the considered models can be found
in their original papers and therefore will not repeated here.
We only point out some noticeable differences between them, which are
relavent to the understanding of the pion cloud effects.

\section{$R_{EM}$ and $R_{SM}$ ratios of the $\Delta$ resonance}

The  deformation of the nucleon and/or $\Delta$ is reflected in the
non-zero values of the
ratios \begin{eqnarray}
R_{EM}&=&\frac{[\bar{\Gamma}]_{E2}}{[\bar{\Gamma}]_{M1}} \\
R_{SM}&=& \frac{[\bar{\Gamma}]_{C2}}{[\bar{\Gamma}]_{M1}},
\end{eqnarray} 
where $[\bar{\Gamma}]_{\alpha}$ denotes 
the $\gamma N \rightarrow \Delta$ transition with 
multipolarity $\alpha = M1$ (magnetic M1), $E2$ (electric E2), 
and $C2$ (Coulomb C2).
As explained, for example, in Ref.\cite{satolee2}, these two ratios at the
resonance energy $W = m_\Delta = 1232 $ MeV can be calculated from
the imaginary ($Im$) parts of the multipole amplitudes 
of pion electroproduction reactions
\begin{eqnarray}
R_{EM} &=& \frac{Im[E^{3/2}_{1^+}]}{Im[M^{3/2}_{1^+}]} \\
R_{SM} &=& \frac{Im[S^{3/2}_{1^+}]}{Im[M^{3/2}_{1^+}]}
\end{eqnarray}
Here the standard notations 
($A^{T}_{\ell\pm}$ with $A =$ $M$, $E$, $S$)\cite{pdg} of the
multipole amplitudes are used
in the right hand sides of the above equations. The  
$R_{EM}$ and  $R_{SM}$ generated from the considered three models
are compared in Fig.1 along with the empirical
values obtained from performing the amplitude  
 analyses of the data from MIT-Bates\cite{bates}, Mainz\cite{mainz},
and Jefferson Laboratory (Jlab)\cite{jlab1,jlab2}. The comparison shown in
Fig.1 was already
given in the publication\cite{jlab2} by the Jlab CLAS collaboration.

We see from Fig.1 that the  $R_{SM}$ (right panel) ratios
predicted by the SL model (solid curve)
and DMT model (dashed curve)
are strikingly different in the $Q^2 \leq 0.5$ (GeV/c)$^2$ region.
However their corresponding predictions of the 
ratio $R_{EM}$ (left panel) are very close in the same low $Q^2$ region. 
As we will see in Figs.3-10, all three models give almost the same
$M_{1^+}$ amplitude at the resonance position $W=1232$ MeV. Thus their
differences seen in Fig.1
can be understood from Fig.2 where the $Q^2$-dependence 
of the predicted ratios $Im[S_{1^+}]/Im[E_{1^+}]$ are compared.
We see that the
SL model prediction at $Q^2=0$ is very close to
the long wavelength limit 
$S_{1^+} \sim E_{1^+}$ ($L_{1^+}=(\omega/\mid \vec{q}\mid) S_{1^+}$) which is
consistent with the results 
given in the papers by Amaldi, 
Fubini, and Furlan\cite{amaldi} and also by Capstick and Karl\cite{capstick}.
On the other hand,
the $Im[S_{1^+}]/Im[E_{1^+}]$ of DMT(MAID2003) model approaches to
2 (3) at $Q^2=0$. Thus their predictions of $R_{SM}$ are 
very different, as seen in the right panel of Fig.1.

As explained in 
section IV of Ref.\cite{satolee2}, the long wave length limit is used 
in the SL model to
define the bare $C2$ transition strength as
$G_C(0) = - [4 m_\Delta^2/(m_\Delta^2 - m_N^2)]G_E(0)$. With $ G_E(0)=+0.025$
determined from fitting the pion photoproduction data, $G_C(0) = -0.238$ 
is fixed in the
SL model $without$ making use of the pion electroproduction data. 
In Fig.1 we see that the $R_{SM}$ data point
at $Q^2 \sim 0.127 $ (GeV/c)$^2$ from 
MIT-Bates\cite{bates} and Mainz\cite{mainz} disagrees with the prediction from 
the SL model, while it agrees with the results generated from
the DMT and MAID models. This difference between the SL 
model and the DMT model marks the large descrepancies between these two
dynamical models in describing the pion cloud effects. As illustrated in 
Fig.10 of Ref.\cite{satolee2}, the pion cloud effect can enhence the 
$\gamma N \rightarrow \Delta$ $C2$ transition( $ \sim Im(S^{3/2}_{1^+}$))
by a factor of about 2 at $Q^2 \sim 0.2 $ (GeV/c)$^2$ and has a very
prounced $Q^2$-dependence. Thus the experimental verifications of the
predictions in the entire low $Q^2$ region given in Fig.1 will lead to
a detailed understanding of the pion cloud effects.

\section{Multipole Amplitudes}

To give more detailed information for analyzing the new data, we compare
in Figs. 3-10 the multipole amplitudes generated from the
SL model (solid curves), DMT model (dashed curves), and MAID2003( dotted
curves). We see that they are in excellent agreement only in the
imaginary part of the M1 multipole ($ Im M_{1^+}$). For other resonant
multipole amplitudes ($E_{1+}$ and $S_{1+}$), the SL model differs
 significantly
from the other two models. These differences can lead to
rather different predictions on
various interference cross sections $\sigma_{LT}$,
$\sigma_{TT}$, and $\sigma_{LT}^\prime$.

For non-resonant
multipole amplitudes in Figs.3-10, the differences between the considered
three models
are very large in some cases. However it is not easy to identify
experimental observables which are most effective in testing these
weaker amplitudes.

Finally, we list in Tables 1-4 
the multipoles amplitudes generated from the SL model at
$ Q^2 =$ 0., 0.05, 0.1, 0.20, 0.5. The corresponding
values from DMT and MAID can be obtained from the web site listed in
Ref.\cite{maid}. More results from the SL model can be obtained
from the authors.

\vspace{2cm}

This work was supported in part by the U.S. Department of Energy,
Office of  Nuclear Physics, under Contract No. W-31-109-ENG-38
and Japan Society for the Promotion
of Science, Grant-in-Aid for Scientific Research (C) 15540275.

\newpage

\begin{table}[htb]
\caption{ The $Q^2$-dependence of the multipole ampliutde for the $\gamma + p
\rightarrow \pi^0 + p$ reaction in unit of $10^{-3}/m_\pi$ at
 $W=1.232$GeV.  $Q^2$ is given in unit of $(GeV/c)^2$.}
\begin{tabular}{lrrrrr} \hline
$Q^2$ &  0 &  0.05 &  0.1 &  0.2 & 0.5\\ \hline
$E_0^+$ & $-0.346+2.105i$ & $0.520+1.961i$ & $1.169+1.786i$ & $1.987+1.454i$ & $2.630+0.837i$\\
$E_1^+$ & $1.056-0.652i$ & $1.067-0.808i$ & $1.003-0.870i$ & $0.848-0.846i$ & $0.539-0.571i$\\
$M_1^+$ & $-2.048+25.008i$ & $-1.857+27.276i$ & $-1.648+27.872i$ & $-1.321+26.989i$ & $-0.826+21.053i$\\
$M_1^-$ & $-2.081+0.311i$ & $-2.508+0.242i$ & $-2.792+0.183i$ & $-3.097+0.103i$ & $-3.118+0.011i$\\
$S_0^+$ & $-1.461+1.784i$ & $-1.351+1.784i$ & $-1.281+1.641i$ & $-1.230+1.332i$ & $-1.131+0.779i$\\
$S_1^+$ & $0.819-0.557i$ & $0.832-0.922i$ & $0.735-1.122i$ & $0.540-1.257i$ & $0.262-1.174i$\\
$S_1^-$ & $0.985+0.407i$ & $1.430+0.460i$ & $1.724+0.449i$ &
 $2.014+0.385i$ & $1.988+0.233i$\\ \hline
\end{tabular}
\end{table}


\begin{table}[htb]
\caption{ The $Q^2$-dependence of the multipole ampliutde for the $\gamma + p
\rightarrow \pi^+ + n$ reaction.}
\begin{tabular}{lrrrrr} \hline
$Q^2$ &  0 &  0.05 &  0.1 &  0.2 & 0.5\\ \hline
$E_0^+$ & $-10.665-0.450i$ & $-10.320-0.601i$ & $-9.718-0.699i$ & $-8.384-0.793i$ & $-5.509-0.779i$\\
$E_1^+$ & $-1.448-0.441i$ & $-1.452-0.552i$ & $-1.358-0.596i$ & $-1.140-0.583i$ & $-0.723-0.394i$\\
$M_1^+$ & $1.144+17.660i$ & $0.715+19.269i$ & $0.376+19.695i$ & $-0.024+19.076i$ & $-0.308+14.884i$\\
$M_1^-$ & $-3.601+0.135i$ & $-2.148+0.156i$ & $-0.946+0.170i$ & $0.610+0.185i$ & $2.161+0.182i$\\
$S_0^+$ & $-8.502-0.128i$ & $-8.555-0.152i$ & $-7.852-0.132i$ & $-6.283-0.065i$ & $-3.485+0.051i$\\
$S_1^+$ & $-1.120-0.378i$ & $-1.113-0.636i$ & $-0.962-0.780i$ & $-0.676-0.879i$ & $-0.288-0.826i$\\
$S_1^-$ & $-7.513-0.040i$ & $-8.714-0.063i$ & $-8.754-0.081i$ &
 $-7.908-0.100i$ & $-5.370-0.106i$\\  \hline
\end{tabular}
\end{table}

\newpage
\begin{table}[htb]
\caption{ The $Q^2$-dependence of the multipole ampliutde for the $\gamma + n
\rightarrow \pi^0 + n$ reaction.}
\begin{tabular}{lrrrrr} \hline
$Q^2$ &  0 &  0.05 &  0.1 &  0.2 & 0.5\\ \hline
$E_0^+$ & $2.188+2.576i$ & $2.478+2.325i$ & $2.666+2.065i$ & $2.811+1.607i$ & $2.518+0.816i$\\
$E_1^+$ & $1.052-0.652i$ & $1.050-0.808i$ & $0.974-0.869i$ & $0.799-0.846i$ & $0.471-0.570i$\\
$M_1^+$ & $-3.374+25.020i$ & $-3.286+27.289i$ & $-3.092+27.886i$ & $-2.690+27.001i$ & $-1.820+21.062i$\\
$M_1^-$ & $-0.905+0.357i$ & $-1.193+0.295i$ & $-1.410+0.238i$ & $-1.685+0.160i$ & $-1.853+0.061i$\\
$S_0^+$ & $0.995+2.241i$ & $1.425+2.301i$ & $1.649+2.186i$ & $1.760+1.888i$ & $1.458+1.260i$\\
$S_1^+$ & $0.834-0.557i$ & $0.850-0.922i$ & $0.755-1.123i$ & $0.559-1.257i$ & $0.267-1.174i$\\
$S_1^-$ & $0.268+0.379i$ & $0.441+0.420i$ & $0.536+0.401i$ &
 $0.582+0.328i$ & $0.416+0.170i$\\  \hline
\end{tabular}
\end{table}

\begin{table}[htb]
\caption{ The $Q^2$-dependence of the multipole ampliutde for the $\gamma + n
\rightarrow \pi^- + p$ reaction.}
\begin{tabular}{lrrrrr} \hline
$Q^2$ &  0 &  0.05 &  0.1 &  0.2 & 0.5\\ \hline
$E_0^+$ & $-14.247-1.116i$ & $-13.089-1.117i$ & $-11.835-1.092i$ & $-9.549-1.010i$ & $-5.350-0.749i$\\
$E_1^+$ & $-1.443-0.441i$ & $-1.428-0.552i$ & $-1.317-0.596i$ & $-1.071-0.583i$ & $-0.627-0.394i$\\
$M_1^+$ & $3.018+17.642i$ & $2.735+19.250i$ & $2.419+19.676i$ & $1.912+19.058i$ & $1.098+14.871i$\\
$M_1^-$ & $-5.265+0.068i$ & $-4.009+0.082i$ & $-2.900+0.092i$ & $-1.388+0.105i$ & $0.372+0.110i$\\
$S_0^+$ & $-11.975-0.774i$ & $-12.481-0.882i$ & $-11.995-0.902i$ & $-10.512-0.852i$ & $-7.146-0.630i$\\
$S_1^+$ & $-1.141-0.378i$ & $-1.139-0.636i$ & $-0.989-0.780i$ & $-0.702-0.879i$ & $-0.296-0.826i$\\
$S_1^-$ & $-6.500+0.001i$ & $-7.316-0.007i$ & $-7.074-0.014i$ &
 $-5.882-0.020i$ & $-3.146-0.017i$\\  \hline
\end{tabular}
\end{table}

\newpage

\begin{figure}[h]
\centering
\includegraphics[width=7cm]{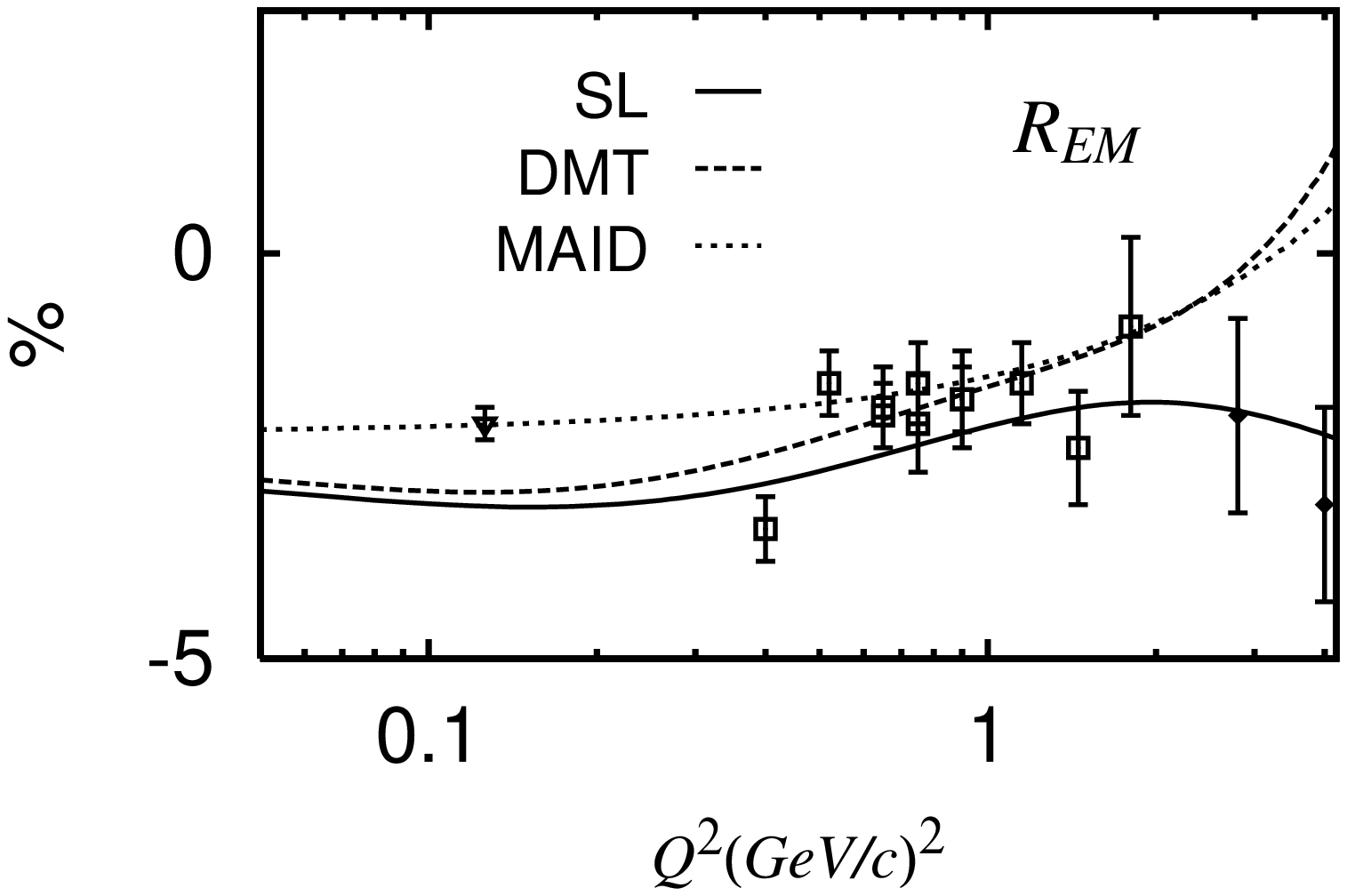}
\includegraphics[width=7cm]{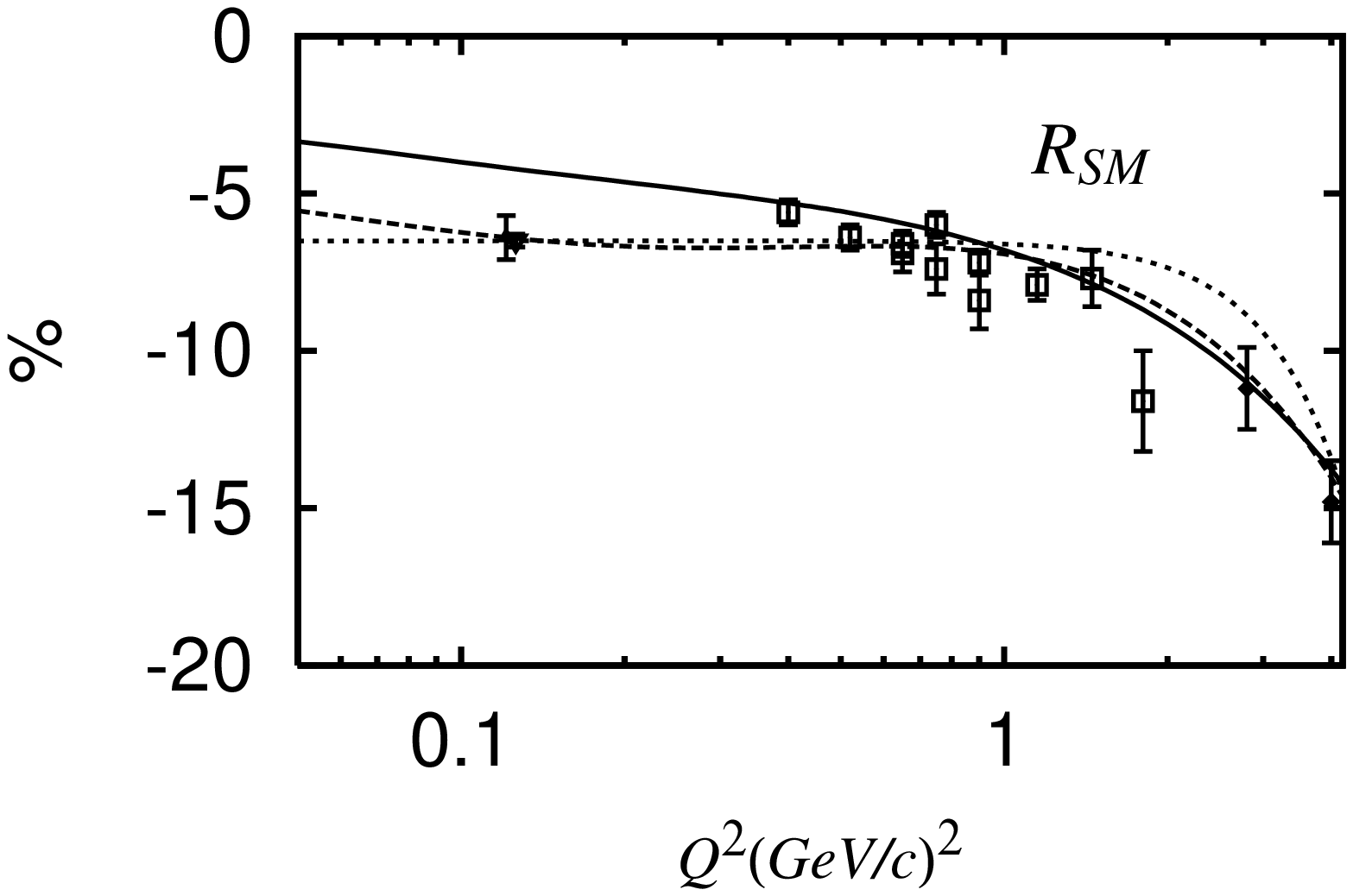}
  \caption{The ratios $R_{EM}$ (left panel) and $R_{SM}$ (right panel) predicted
by the SL model[1,2] (solid curves), DMT model[3,4] (dashed curves), and
MAID2003[5] (dotted curves) are compared with the empirical values from
MIT-Bates[7], Mainz[8], and Jefferson Laboratory[9,10].}
\end{figure}

\begin{figure}[h]
\centering
\includegraphics[width=7cm]{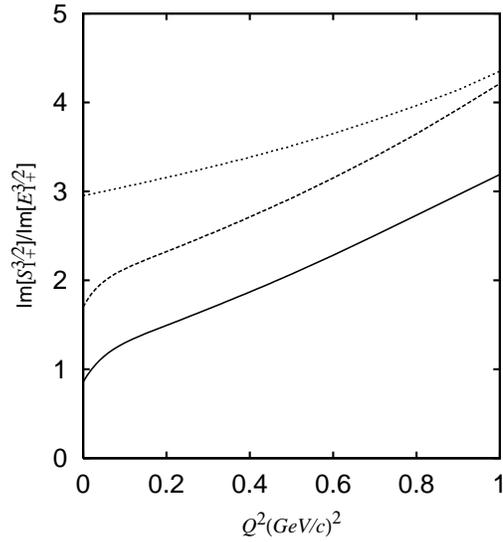}
 \caption{The ratios $Im(S^{3/2}_{1+})/Im(E^{3/2}_{1+})$ predicted
by the SL model[1,2] (solid curves), DMT model[3,4] (dashed curves), and
MAID2003[5] (dotted curves) are compared.}
\end{figure}

\begin{figure}[h]
\centering
\includegraphics[width=14cm]{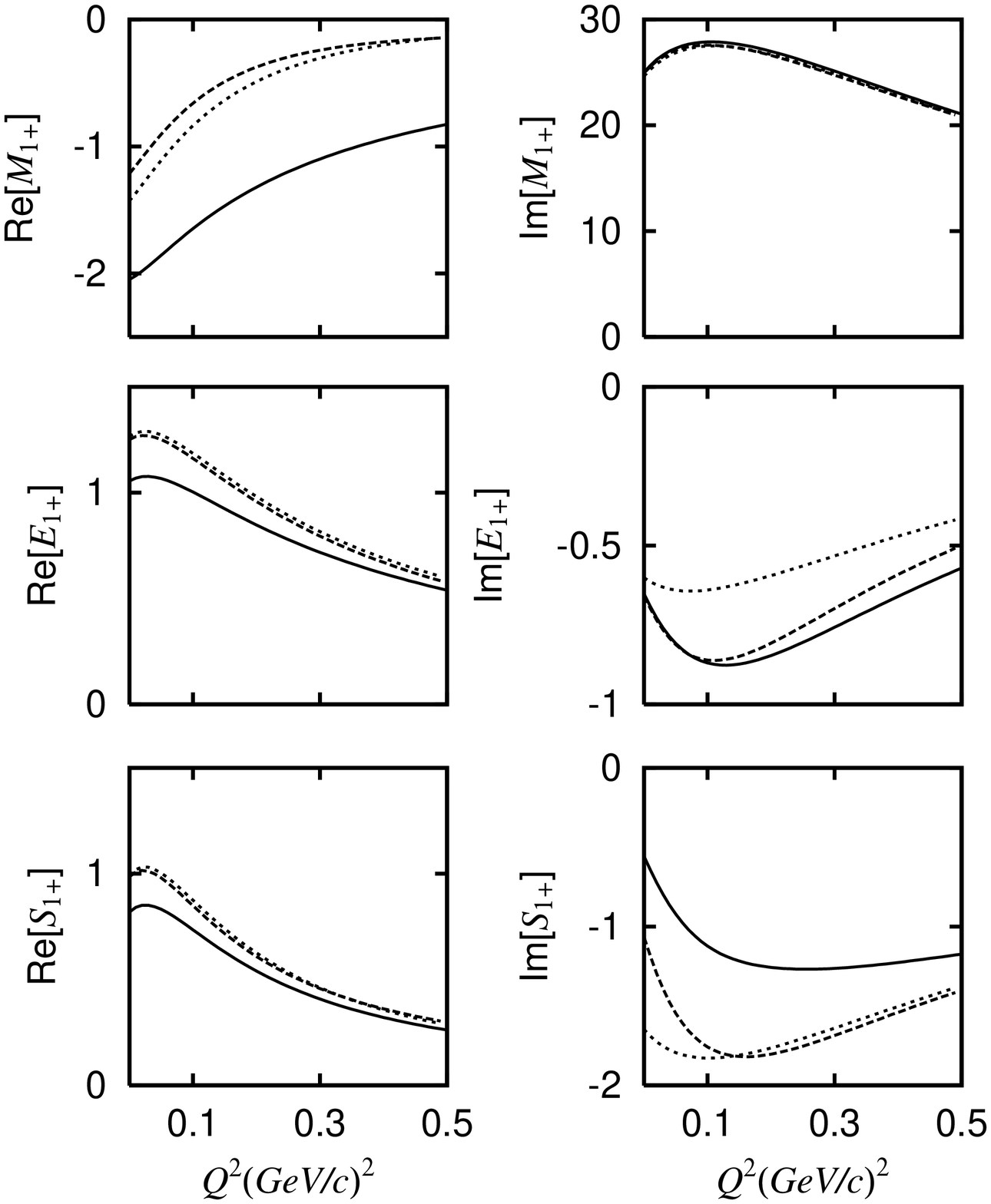}
 \caption{Resonant multipole amplitudes of the $\gamma + p \rightarrow \pi^0 + p$ 
reaction at $W=1232$ MeV.}
\end{figure}

\begin{figure}[h]
\centering
\includegraphics[width=14cm]{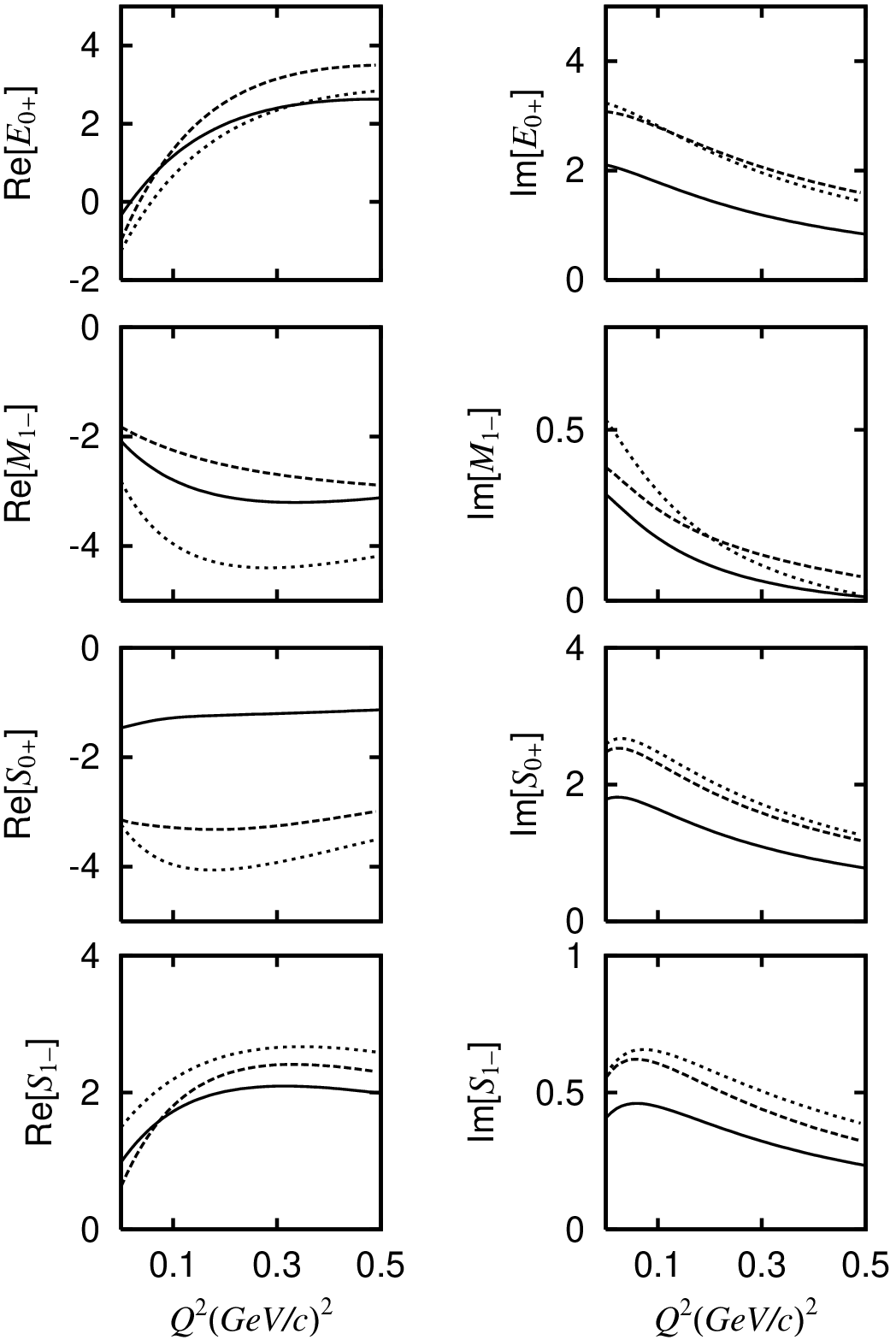}
 \caption{Non-resonant multipole amplitudes of the $\gamma^* p \rightarrow \pi^0  p$
reaction at $W=1232$ MeV. }
\end{figure}

\begin{figure}[h]
\centering
\includegraphics[width=14cm]{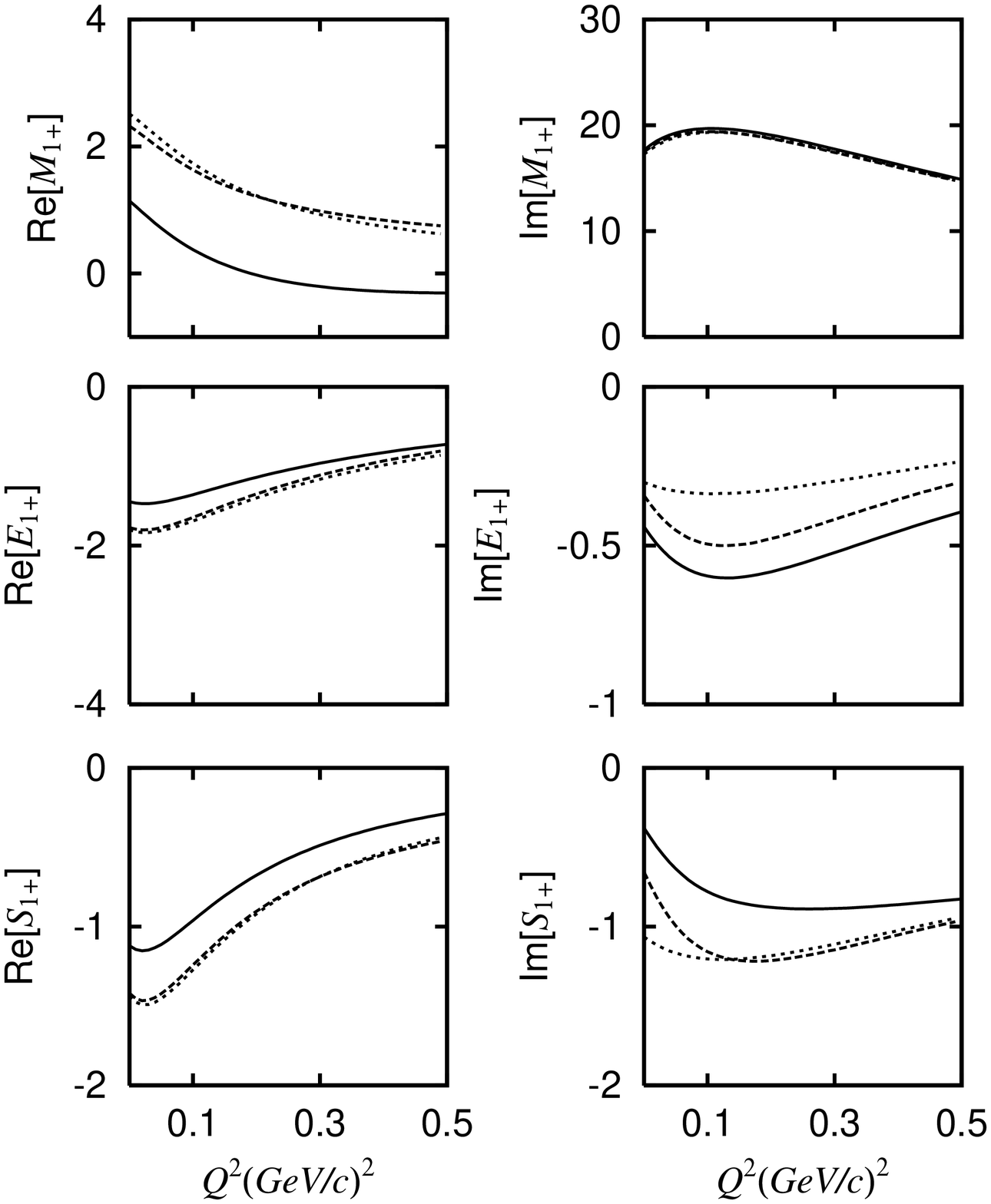}
 \caption{Resonant multipole amplitudes of the $\gamma^*  p \rightarrow \pi^+  n$
reaction at $W=1232$ MeV. }
\end{figure}

\begin{figure}[h]
\centering
\includegraphics[width=14cm]{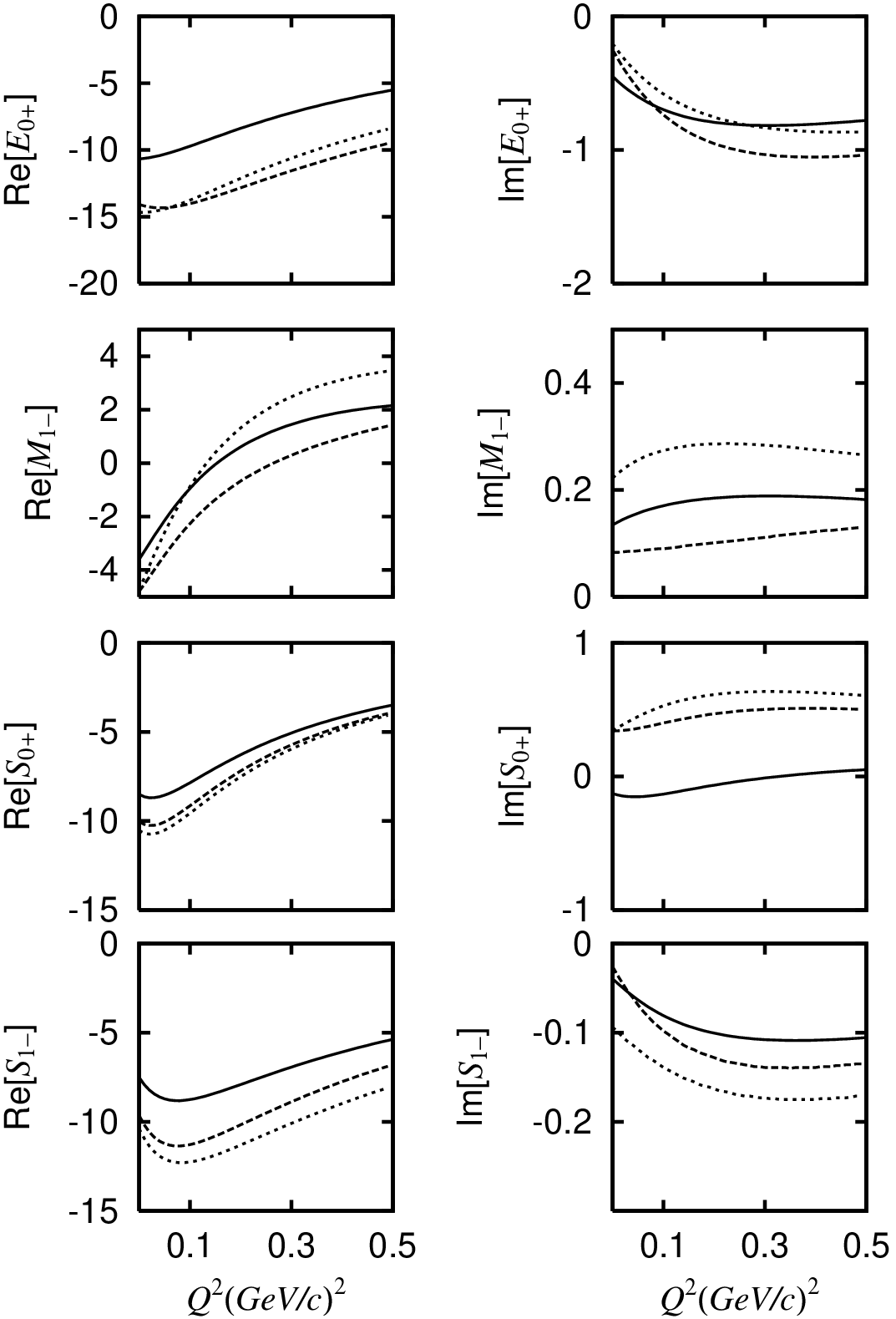}
 \caption{Non-resonant multipole amplitudes of the $\gamma^*  p \rightarrow \pi^+  n$
reaction at $W=1232$ MeV. }
\end{figure}

\begin{figure}[h]
\centering
\includegraphics[width=14cm]{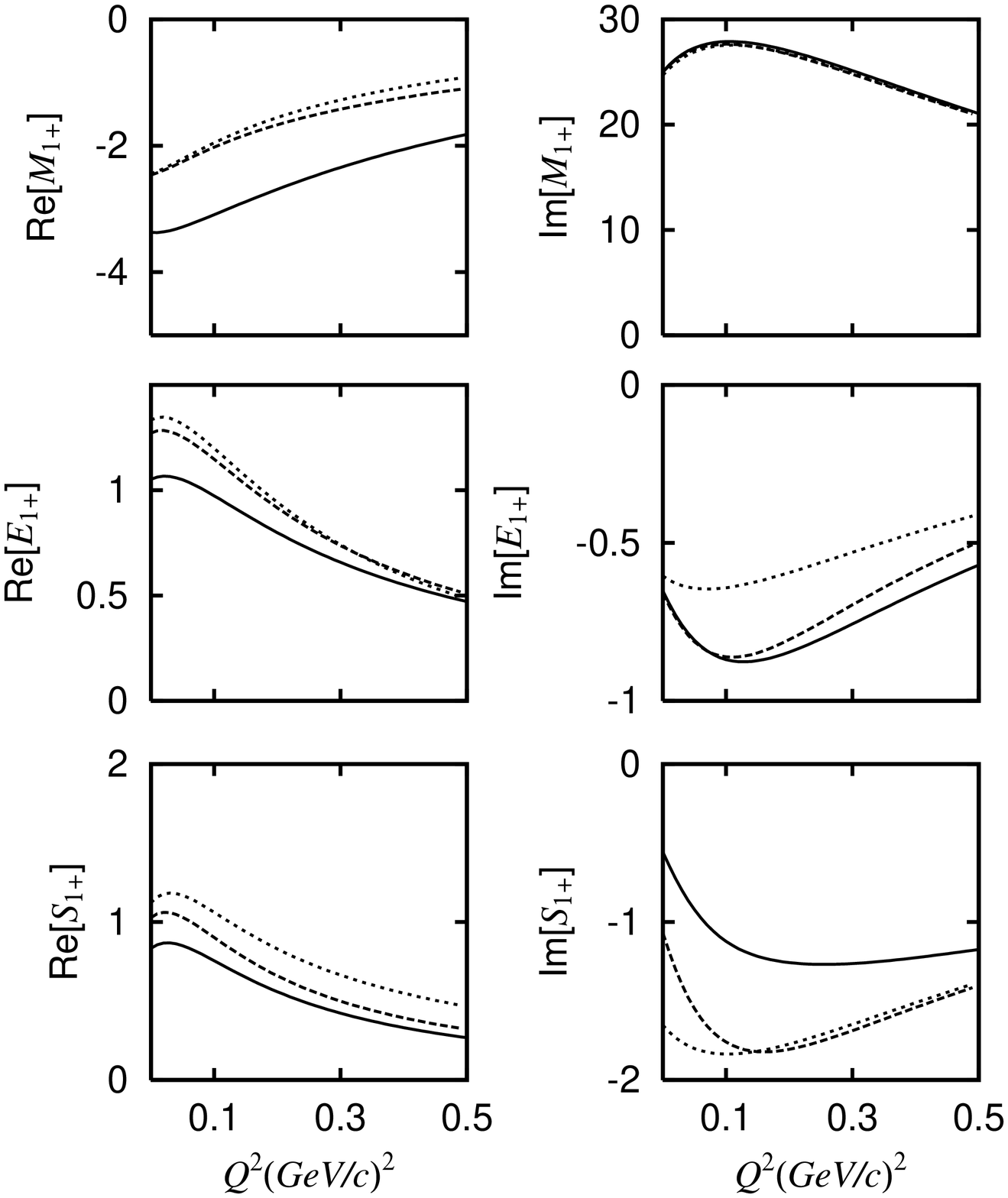}
 \caption{Resonant multipole amplitudes of the $\gamma^*  n \rightarrow \pi^0  n$
reaction at $W=1232$ MeV. }
\end{figure}

\begin{figure}[h]
\centering
\includegraphics[width=14cm]{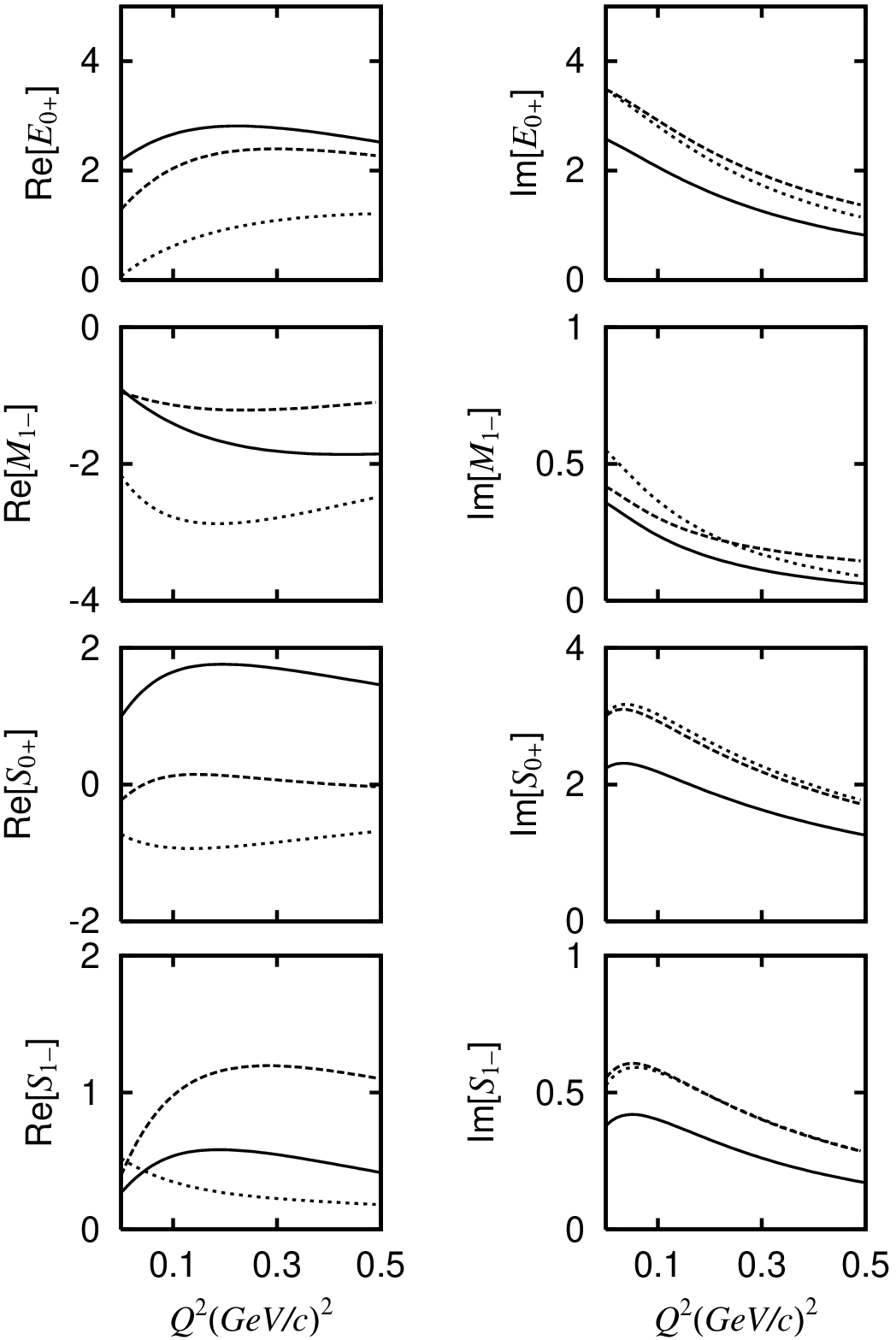}
 \caption{Non-resonant multipole amplitudes of the $\gamma^*  n \rightarrow \pi^0  n$
reaction at $W=1232$ MeV. }
\end{figure}

\begin{figure}[h]
\centering
\includegraphics[width=14cm]{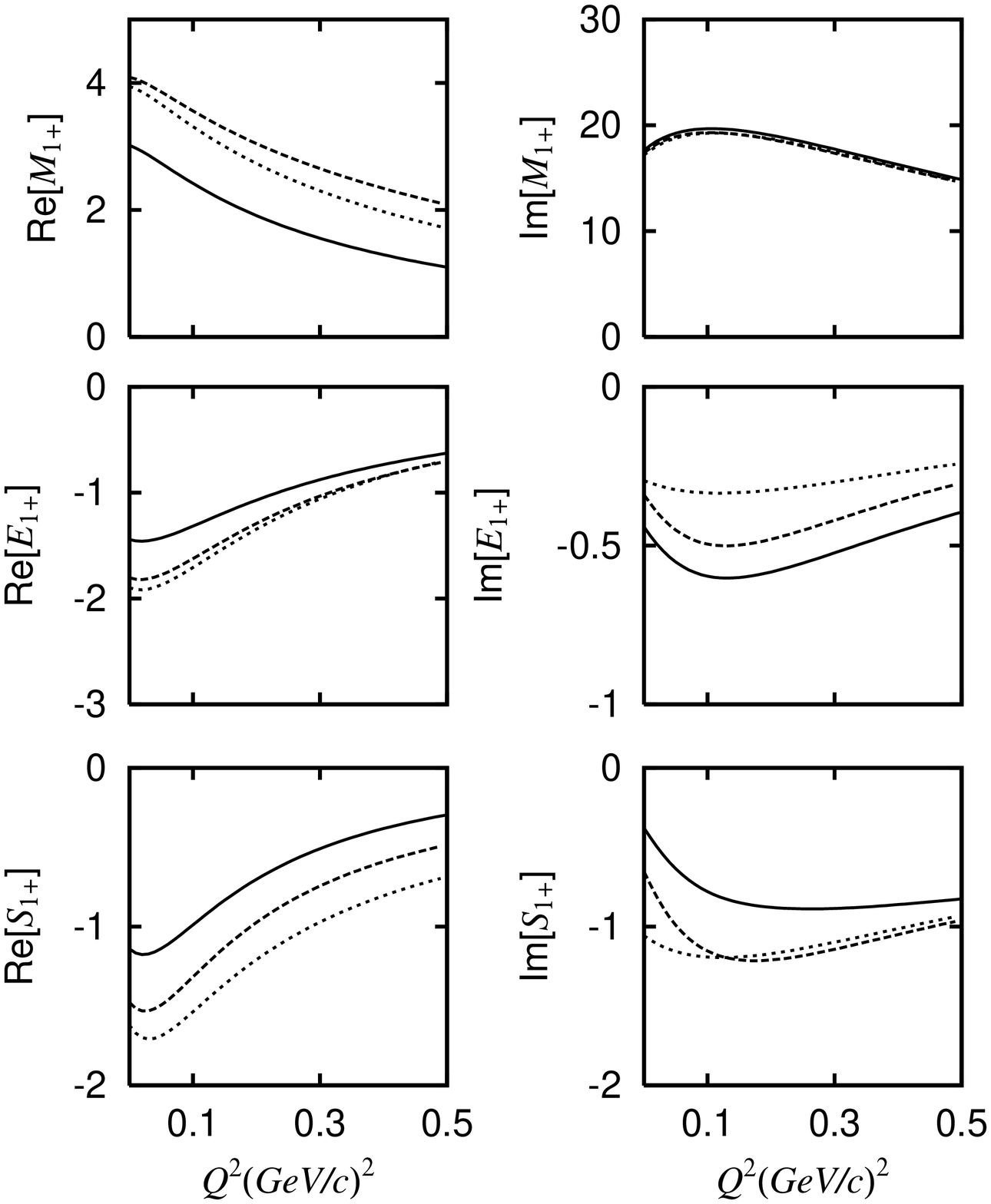}
 \caption{Resonant multipole amplitudes of the $\gamma^*  n 
\rightarrow \pi^-  p$
reaction at $W=1232$ MeV. }
\end{figure}
\begin{figure}[h]
\centering
\includegraphics[width=14cm]{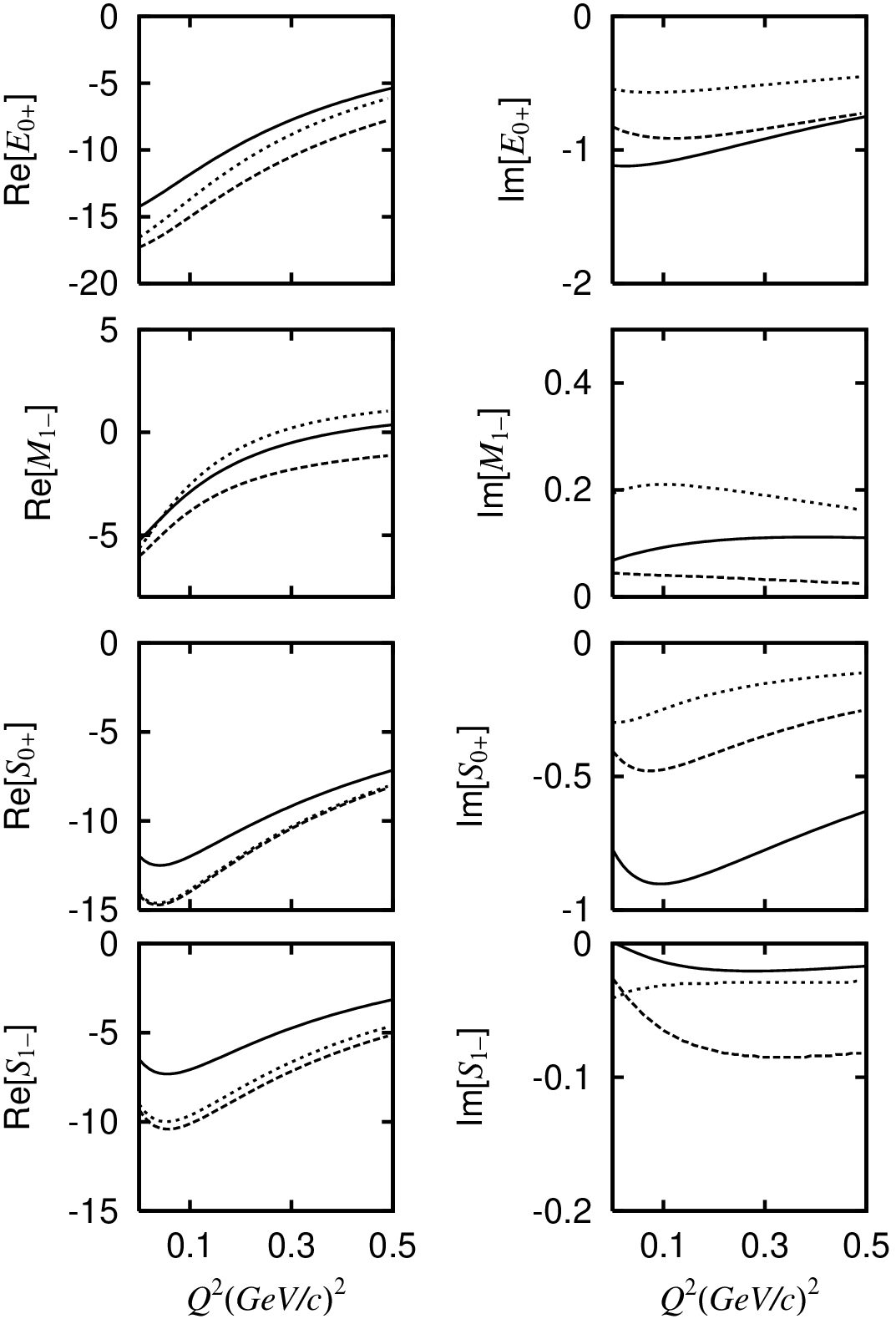}
 \caption{Non-resonant multipole amplitudes of the $\gamma^*  n 
\rightarrow \pi^-  p$
reaction at $W=1232$ MeV. }
\end{figure}


\newpage
\begin{references}
\bibitem{satolee1}
T.~Sato and T.-S.~H. Lee, Phys. Rev. C {\bf 54}, 2660 (1996).

\bibitem{satolee2}
T.~Sato and T.-S.~H. Lee, Phys. Rev. C {\bf 63}, 055201 (2001).

\bibitem{dmt1}
S. Kamalov and S. N. Yang, Phys. Rev. Lett {\bf 83}, 4494 (1999)

\bibitem{dmt2}
S. Kamalov et. al., Phys. Rev. {\bf C64}, 032201 (2001)

\bibitem{maid}
D. Dreschel et al., Nucl. Phys. {\bf A645}, 145 (1999);
http://www.kph.uni-mainz.de/T/maid/maid.html; Here we show their results
of 2003.

\bibitem{pdg}
Particle Data Group, D. E. Groom et al., Eur. Phys. J. {\bf C15}, 1 (2000).

\bibitem{bates}
C. Mertz et al., Phys. Rev. Lett {\bf 86}, 2963 (2001).

\bibitem{mainz}
R. Beck et al., Phys. Rev. {\bf C61}, 035204 (2000); T. Pospischil et al.,
Phys. Rev. Lett. {\bf 86}, 2959 (2001).

\bibitem{jlab1}
V.V. Frolov et al., Phys. Rev. Lett. {\bf 82}, 45 (1999)

\bibitem{jlab2}
K. Joo et. al., Phys. Rev. Lett. {\bf 88}, 12201 (2002) 

\bibitem{amaldi}
See Eq.(C13) of E. Amaldi, S. Fubini, and G. Furlan, in Springer Tracts in Mordern Physics, 
Vol. 83, edited by G. Hohler (Springer, Berline, 1979), p.1.

\bibitem{capstick}
See second paragraph of Sec. III of S. Capstick and G. Karl, 
Phys. Rev. {\bf D 41}, 2767 (1990).
\end{references}
\end{document}